\def\aa{{A\&A}}

\def\aj{{AJ}}

\def\apj{{ApJ}}

\def\etal{{et al.\,}}

\documentclass[cup5b]{caps}
\usepackage{graphicx}
\usepackage{amssymb}
\usepackage{ociwsymp4e}  

\HeadText{K. A. Venn et al.}

\begin{document}

\pagenumbering{arabic}

\author[]{K. A. Venn$^{1}$, E. Tolstoy$^{2}$, A. Kaufer$^{3}$, 
          R. P. Kudritzki$^{4}$ \\
(1) Macalester College and University of Minnesota, MN, USA\\
(2) Kapteyn Institute, University of Groningen, Netherlands\\
(3) European Southern Observatory, Santiago, Chile\\
(4) Institute for Astronomy, University of Hawaii, HI, USA\\ 
}

\chapter{Stellar Abundances in Dwarf Irregular Galaxies}

\begin{abstract}
Dwarf irregular galaxies appear to have undergone very slow 
chemical evolution since they have low nebular abundances,
but have had ongoing star formation over the past 15 Gyr.   
They are too distant for red giant abundance analyses to 
examine the details of their chemical evolution, 
however the isolated, bright blue supergiants
do allow us to determine their present-day iron abundances to
compare with both stellar and nebular $\alpha$-element results. 
The [$\alpha$/Fe] ratios in four Local Group dwarf irregular
galaxies (NGC\,6822, WLM, Sextans~A, and GR~8) all appear to
have solar ratios regardless of the differences in their 
metallicities and star formation histories.    Surprisingly,
WLM's stellar metallicity is three times higher than the nebular
oxygen abundance.   We compare the [$\alpha$/Fe] ratios in the
dwarf irregulars to those from recent analyses of red giant
branch stars in dwarf spheroidal galaxies, and also to damped
Ly$\alpha$ systems, and discuss these in the context of model
predictions.

\end{abstract}

\section{Introduction}

A examination of the distribution of galaxies in the Local Group
(e.g., Figure 3 by Grebel 1999) clearly shows the density-morphology
relationship for galaxies.   The dwarf spheroidals are all located
near the large mass concentrations of the large spirals, with the
dwarf irregulars located at greater distances.   One of the most 
useful ways to study the star formation and chemical evolution of
a galaxy is to examine the elemental abundances and abundance ratios
in its stars.  In the Magellanic Clouds and the nearby dwarf spheroidals,
it is possible to determine the abundance of a wide range of elements
from the red giant branch (RGB) stars, and since these stars have a range 
of metallicities and ages, then it is possible to build up a picture of
the chemical evolution of those galaxies; for more details see the 
contributions in this proceedings by Hill, Shetrone, and Smith.

The dwarf irregular galaxies appear to have undergone very little and
very slow chemical evolution since they have very low nebular abundances
(down to 1/20 solar, e.g., Skillman \etal 1989a).   
If dwarf galaxies are the corner stones for galaxy formation, as in the
hierarchical structure simulations in the cold dark matter (CDM) scenarios,
then the dwarf irregulars may be the cleanest remnants of the proto-galactic 
fragments from the high redshift Universe. 
The dwarf irregular galaxies are too distant for detailed chemical
abundance analyses of their RGB stars.   However, there are relatively
isolated, blue supergiants in these galaxies that are bright enough for 
detailed spectroscopy using the efficient spectrographs on the new 
8 to 10-meter class telescopes.   These stars are all young, thus we
cannot examine the build-up of elements from these systems as with the
RGB stars, but we can examine the abundance of elements not available
from nebular analyses (e.g., Fe) to examine the integrated chemical
evolution in these systems to the present epoch. 

\section{Stellar Abundances in NGC\,6822}
 
\begin{figure}
\centering
\includegraphics[width=10cm,angle=0]{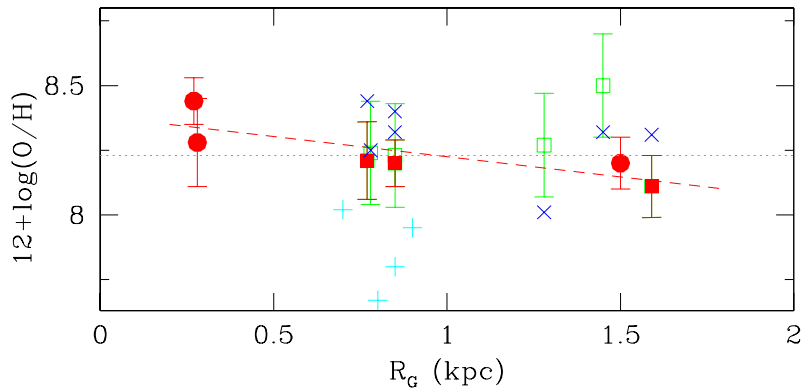}
\caption{Oxygen abundances versus galactocentric distance in NGC\,6822.  
The results from three A-type supergiants (Venn \etal 2001) 
are shown by filled circles, and for three nebulae where the 
OIII 4363 line was detected (Pagel \etal 1980) by filled squares.   
A least squares linear fit to these six (high quality) data points yields
a slope that suggests an abundance gradient of $-$0.1~dex/kpc, which is
the same gradient found in the Galaxy over a much larger scale
(Smartt \& Rolleston 1997).  Hollow squares show other nebular data
(Pagel \etal 1980); x's note the nebular results when the Pilyugin (2001) 
R23 calibration is used; +'s note results for two nebulae 
(Hubble V and X) from other authors.  
All of the data taken together (of various qualities)
is consistent with no slope in the oxygen abundance.  
}
\label{kvenn1}
\end{figure}

The closest, relatively isolated dwarf irregular galaxy is NGC\,6822,
with D$_{6822}$ $\sim$ 0.5 Mpc.
A detailed color-magnitude diagram (CMD) by Gallart \etal (1996a,b)
shows that the brightest blue supergiants in this galaxy are nearly
2 magnitudes brighter than the tip of the red giant branch.  From
Keck-HIRES and VLT-UVES spectroscopy, we showed that two stars
located near the center of NGC\,6822 have oxygen and iron abundances
that are in excellent agreement with the low nebular oxygen abundances
(Venn \etal 2001)\footnote{For comparison, solar abundances are from
Grevesse \& Sauval (1998), except for oxygen, which is adopted as
12+log(O/H) = 8.66 from Asplund (2003, consistent with Allende-Prieto
\etal 2001).}; i.e., [O/H] = $-$0.5 and [Fe/H] = $-$0.5. 
Preliminary analysis of a third star is in agreement with these results
as well, though this star is located in the outer disk of the galaxy.

An interesting sidelight to these abundances in NGC\,6822 is that 
the stellar and nebular oxygen results may show evidence for an
abundance gradient.  A marginal detection of a $-$0.1 dex/kpc
gradient can be seen from a least-squares linear fit to the stellar
data with the nebular abundances derived only from HII regions where
the OIII 4363 line is observed; see Fig.~\ref{kvenn1} .   
This gradient is the same as that
seen in the Galactic disk over approximately 20~kpc, or $\sim$10x
larger distance than NGC\,6822 (e.g., Smartt \& Rolleston 1997).    
Small scale abundance fluctuations,
such as a gradient, are not expected in dwarf galaxies where it is 
thought that the mixing processes occur on short timescales to throughly
mix the interstellar medium.   Confirmation of an abundance gradient
would suggest much longer mixing timescales, {\it in general}, for newly
processed material from SNe~II and massive stars. 

\section{Stellar Abundances in WLM }

\begin{figure}[t]
\centering
\includegraphics[width=10cm,angle=0]{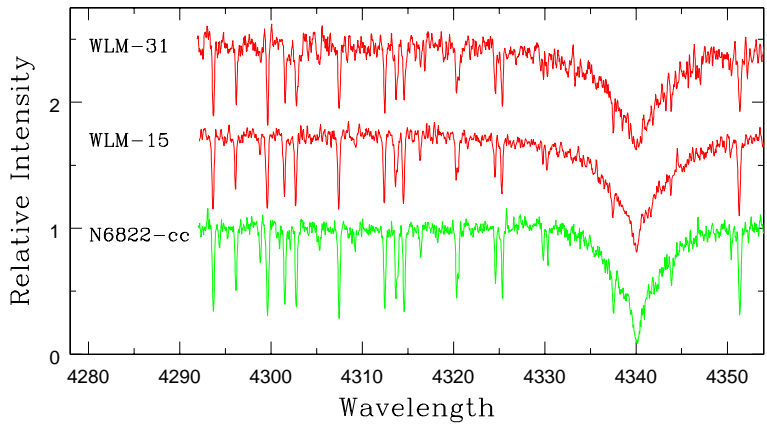}
\caption{Comparison of VLT+UVES spectra around H$\gamma$ 
for two stars in WLM and one in NGC 6822.   All three stars
have very similar atmospheric parameters and metallicities.} 
\label{kvenn3}
\end{figure}

The dwarf irregular galaxy WLM is more distant than NGC\,6822 at
D$_{WLM}$ $\sim$ 1~Mpc, but it also has lower foreground extinction
putting the blue supergiants at similar magnitudes in both of these
galaxies.   The nebular abundances in WLM are much lower than in
NGC\,6822 though.   Hodge \& Miller (1995) and Skillman \etal (1989a) 
have analysed three HII regions (totalled); the OIII 4363 is detected 
and the results are in good agreement with one another at [O/H] = $-$0.9.
Thus, this galaxy is about four times {\it more metal-poor than the SMC},
making it one of the most metal-poor galaxies in the Local Group and
suggesting that its chemical evolution has been quite slow. 

Spectra of two A-type supergiants were obtained with the VLT+UVES
and analysed using standard techniques (Venn \etal 2003).   The stars
have [Fe/H] = $-$0.4, which is nearly 3 $\sigma$ higher than the nebular
oxygen abundance.   In fact, this metallicity is in better agreement 
with the results from NGC\,6822, which can be seen directly by comparing
a portion of the spectra from the two stars in WLM to a star with similar
parameters in NGC\,6822 (see Fig.~ref{kvenn3}).
The WLM stellar oxygen abundance is even higher, [O/H] = $-$0.2,
but this was determined directly from only one star; spectrum synthesis
for this one star is shown in Fig.~ref{kvenn2}.
The radial velocities and stellar parameters do support that these stars
are members of WLM.   The question then is how can these young stars 
be more metal-rich than the nebulae? 

\begin{figure}
\centering
\includegraphics[width=10cm,angle=0]{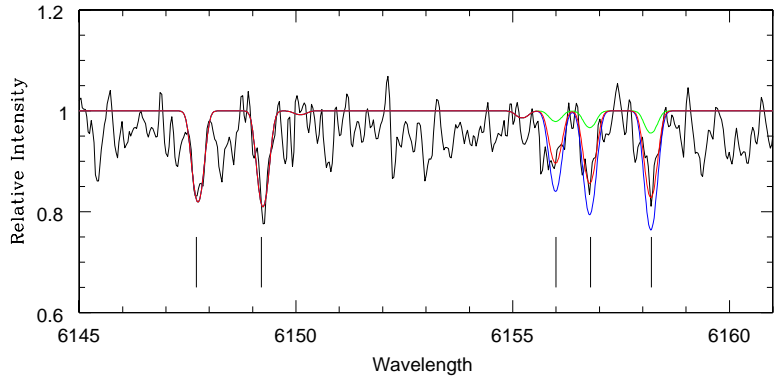}
\caption{Spectrum synthesis for two Fe II lines, 
$\lambda$6147 and $\lambda$6149, and the three 
OI lines near $\lambda$6158 in WLM-15.   
The best fits to the data are for
[Fe/H] = $-$0.4 and [O/H] = $-$0.2.   The synthesis for an
oxygen abundance that is 0.3~dex larger is shown (too large), 
as well as the nebular oxygen abundance (too small).} 
\label{kvenn2}
\end{figure}

At present, the most likely scenario is that there are large scale
abundance variations in WLM.   The two stars are located on the
south-east side of the galaxy, whereas the nebular are in the
central-west portions.   If this is an edge on disk galaxy as
suggested by its HI rotation curve (J.\, Cannon, 
{\it private communication}), then their positions within WLM 
could be even further apart.   However, this galaxy is also significantly
smaller than NGC\,6822; it has about 1/10 the mass and radius.   It would
be surprising to find such large abundance variation in WLM and not have
seen them in NGC\,6822. 
Other possibilities seem even more unlikely though, e.g, dilution through
the recent merger of a large HI cloud (merger would have needed to occur
{\it after} the stars formed only 10 Myr ago), or variations in the
interstellar gas-to-dust ratio (such as large amounts of oxygen locked
in dust grains).   
It is not clear yet what the difference in the stellar and nebular
abundances are telling us. Is there something peculiar about WLM, or
is this result telling us something general about chemical evolution,
nucleosynthesis, or abundance measurements? 

\section{The All-Important [$\alpha$/Fe] Ratio}

The evolution of the chemical abundances in a galaxy is intimately 
linked to its star formation history.   Different elements are produced
during the evolution of stars of different masses, and over a range of
timescales.   If the star formation in a galaxy proceeds by a series of
bursts, rather than smooth, approximately constant star formation, then
this should lead to clear differences in the evolution of the chemical
abundances.   One ratio of particular importance is the $\alpha$/Fe ratio,
typically characterized by [O/Fe].
Oxygen is produced primarily in the
high-mass stars of negligible lifetimes and ejected by SNe II, while iron
is produced in both SNe II and SNe Ia.   Stars that form shortly after
the interstellar medium has been enriched by SNe II may have enriched
[$\alpha$/Fe] ratios, while those that form sometime after the SNe Ia
contribute will have lower [$\alpha$/Fe].  For more details, see the
contributions in this proceedings by Matteucci and Chiappini.  

Since bursts of star formation allow SNe II to contribute $\alpha$-elements,
and long quiescent periods allow SNe Ia to contribute iron, then the 
total [$\alpha$/Fe] ratios in a galaxy should be able to vary over time. 
Gilmore \& Wyse (1991) wrote ``{\it There is nothing special or universal
about solar element ratios, and one should not expect} [solar abundances]
{\it in any other environment which has had a different star formation 
history}''. 
Nevertheless, NGC\,6822 has had a significantly different star formation
history from the solar neighbourhood (see Gallart \etal 1996a,b), and yet 
its young stars and nebulae have the solar ratio of [O/Fe].  The same
is true for the Magellanic Clouds (e.g., Hill 1997, 1999; Hill \etal 1995;
Venn 1999; Rolleston \etal 2003).   A comparison of the
star formation histories for the Magellanic Clouds (e.g., Pagel \&
Tautvaisiene 1998) and NGC\,6822 (Gallart \etal 1996a,b) suggests that
NGC\,6822 has had almost the opposite star formation history
(more old and intermediate-aged stars, few stars forming in the past
5~Gyr, until a recent burst within the past 1~Gyr).
How do such different systems result in similar ratios, and why
are they all near solar? 

\begin{figure}[t]
\centering
\includegraphics[width=10cm,angle=0]{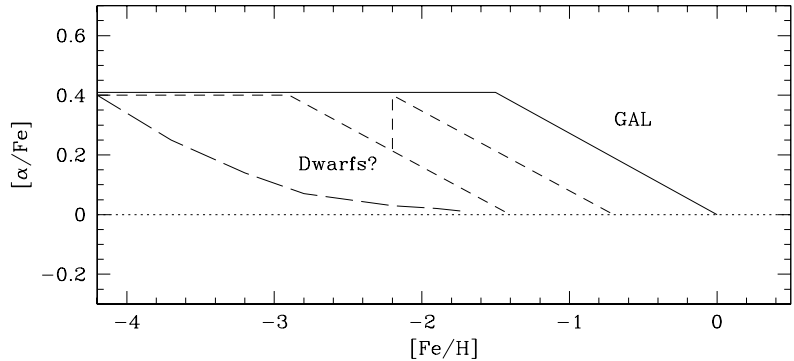}
\caption{Cartoon of [$\alpha$/Fe] versus [Fe/H] to illustrate some of
 the possibilities from different star formation histories 
 and/or different star formation rates.   The solid line represents
 the trend in abundance ratios for Galactic stars.   The short dashed
 lines represent a smaller star formation event where the same peak
 SNe II [$\alpha$/Fe] ratio is reached after an initial star formation
 epoch but at a lower metallicity such that the SNe Ia can contribute
 significant amounts of iron sooner in the chemical evolution.   At
 some later time, a new burst of star formation might increase 
 [$\alpha$/Fe] back to the peak SNe II ratio.   Alternatively, 
 a very slow star formation rate such as in dwarf galaxies may allow
 the SNe Ia to contribute at very low metallicities, such that 
 [$\alpha$/Fe] approaches the solar ratio sooner.   } 
\label{kvenn4}
\end{figure}

Matteucci (2002) had discussed the components that determine
galactic evolution and notes that while the star formation history
is important in determining absolute abundances at any given time,
it is less important when determining abundance ratios.   For ratios,
the stellar lifetimes, IMF, and nucleosynthesis yields are critical.
Thus, galaxies with low star formation rates either in bursts or
continuous should show a short plateau in [$\alpha$/Fe] at low
metallcities (irregulars and spirals), whereas galaxies with high star 
formation rates early in thier lifetimes will have longer plateaus 
(bulges, and ellipticals). 
A cartoon sketch based on Matteucci's (2002) Figure~1 is shown in 
Fig.~\ref{kvenn4}, where we have also added the potential effects of a 
strong intermediate-aged burst following Gilmore \& Wyse (1991).
That all chemical evolution models for all systems tend towards the
solar ratio at the present epoch is interesting, and suggests that
stellar lifetimes, nucleosynthetic yields, and the IMF, are universal;
possibly also that mixing is efficient throughout these galaxies 
on a relatively short timescale ($\sim$1 Gyr). 
It is remarkable that any star formation that occurs after the 
initial epoch seems to have little influence on the present-day
ratio, although this may also be related to the metallicity having
already increased to the point where the addition of new atoms 
simply doesn't affect the total abundance very much.

\begin{figure}[t]
\centering
\includegraphics[width=10cm,angle=0]{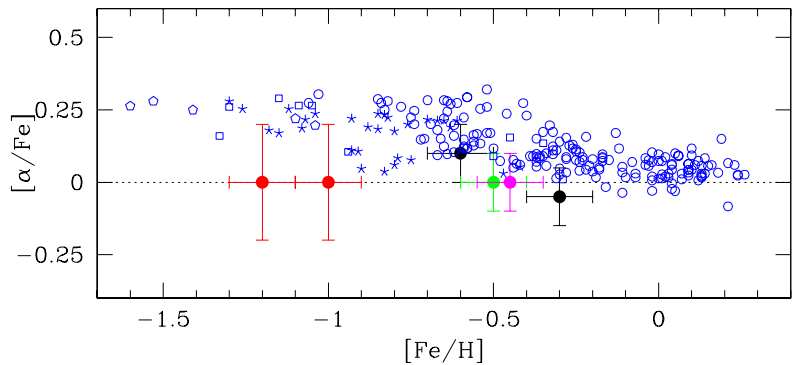}
\caption{[$\alpha$/Fe] ratios for young stars in the dwarf irregular
galaxies and the Magellanic Clouds, compared to Galactic stars.
Filled symbols with errorbars are the mean of the young supergiant
abundances, in order of increasing metallicity, in Sextans A, GR8,
SMC, NGC\,6822, WLM, and the LMC (references in text).  Two $\sigma$ 
errors are plotted for the Sextans~A and GR~8 stars because that 
work is still in progress.  Hollow 
circles represent Galactic data (Edvardsson \etal 1993; Nissen 1997;
Stephens 1999; Gratton \& Sneden 1988, 1991, 1994; McWilliam \etal 1995). 
}
\label{kvenn5}
\end{figure}

This scenario might also help to explain why our preliminary results 
from two more metal-poor dwarf irregular galaxies, Sextans~A and GR~8, 
also have solar-like [$\alpha$/Fe] ratios; see Fig.~\ref{kvenn5}.  
Unlike WLM, our preliminary stellar abundances in individual 
A-type supergiants in Sextans~A and GR~8 from VLT+UVES are consistent 
with the low metallicity determined from their HII regions   
(Van Zee, Skillman \& Hanes 1999; Skillman \etal 1989b)
for both oxygen {\it and} iron.
The abundance ratios for WLM and NGC\,6822 are taken from the 
work discussed above.
The abundance ratios for the young stars in the LMC and SMC are 
taken from the literature for B-dwarfs to red giants, which are
usually in excellent agreement (e.g., Rolleston \etal 2003; 
Venn 1999; Hill \etal 1997, 1999; Luck \etal 1998).

\begin{figure}
\centering
\includegraphics[width=10cm,angle=0]{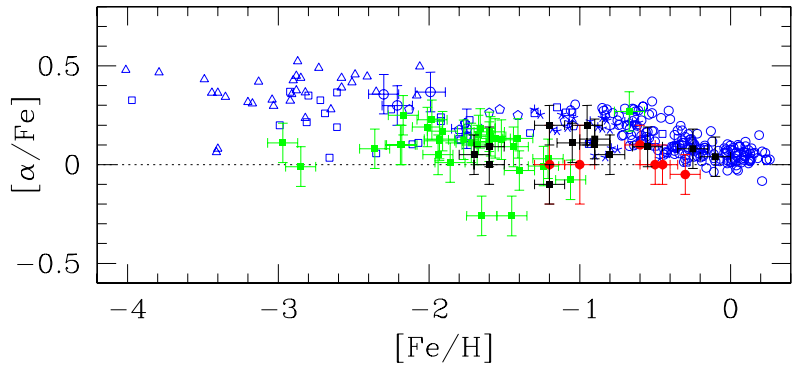}
\caption{[$\alpha$/Fe] ratios for stars in the dwarf irregulars (red
circles), dwarf spheroidals (green squares), and DLAs (black triangles), 
compared to Galactic stars.  
Clearly the [$\alpha$/Fe] ratio does not reach the peak SNe II 
contribution in any of these systems.   But also, the stars in
these systems rarely show subsolar [$\alpha$/Fe]. 
See text and Fig.~\ref{kvenn5} for references.}
\label{kvenn6}
\end{figure}

We also plot the [$\alpha$/Fe] ratio in dwarf spheroidal galaxies 
(Shetrone \etal 2001, 2003) and damped Ly$\alpha$ systems 
(DLAs, Ledoux \etal 2002); see Fig.~\ref{kvenn6}.   
Again, we see that the ratios are lower 
than those in the Galactic metal-poor stars in all of these systems.
Since the star formation histories are signficantly different 
between each of these systems, e.g., some dwarf spheroidals seem 
to have formed all of their stars $>$10 Gyr ago (Sculptor), while 
others have undergone distinct burst of star formation at 
intermediate ages (Carina), and the dwarf irregulars are still
undergoing strong star formation events, then it is impressive
that they all have similar low [$\alpha$/Fe] ratios.

A few systems may show the plateau in the [$\alpha$/Fe] ratios
explected in their metal-poor stars when individual elements
are examined; in particular the LMC and Sculptor may show the
plateau in [Mg/Fe] in their oldest stars (see the conference
contributions by Hill, Shetrone, and Smith in this proceedings).. 
That [Mg/Fe] may differ from the mean [$\alpha$/Fe] suggests
the formation of Mg should be considered separately from heavier
$\alpha$-elements such as Ca and Ti.
Also, the fact that most stars in the dwarf spheroidals show  
$\alpha$-element ratios below the plateau
suggests that they all have SNe Ia contributions,
thus star formation had to be longer than the timescale for SNe Ia
enrichment and mixing into the interstellar medium.
Finally, two stars in the dwarf spheroidals do appear to have
[$\alpha$/Fe] that is less than solar.   These are Carina-M3 
(Shetrone \etal 2003) and Sextans-58 (Shetrone \etal 2001).    
Tolstoy \etal (2003) estimate the ages for these stars from their
red giant position on the CMD; Carina-M3 is 13 $\pm$3 Gyr and
Sextans-58 is $\sim$6 Gyr.   While ages are not very well constrained,
Sextans-58 is the youngest star analysed in that galaxy which may 
suggest its chemical evolution has pushed its abundance ratios below
the solar value.   Carina-M3 is not young though, it is one of the
oldest stars analysed in that galaxy.   However, Carina has had a
very complex star formation history with at least three distinct 
star formation epochs.   Perhaps the chemical evolution of this
galaxy evolved to subsolar ratios before the intermediate-aged
burst of star formation raised [$\alpha$/Fe] again.

\section{Acknowledgements}
Many thanks to Andy McWilliam and Michael Rauch for organizing 
such a stimulating and interesting conference, also for their 
generous flexibility in scheduling.  KAV would like to
thank the NSF for support through a CAREER award, AST-9984073.
Also thanks to the Institute of Astronomy, University of Cambridge,
for support and hospitality while much of this work has been done.

\end{document}